\documentclass[11pt,twoside]{article}
\usepackage{jltp}
\usepackage{amssymb}
\title{From quantum disorder to quantum chaos \thanks{dedicated
to Peter W\"olfle on the occasion of his 60th birthday}}

\author{I.~V.~Gornyi$^{1,2,\dagger}$ and
A.~D.~Mirlin$^{1,2,\#}$}
\address{$^1$Institut
f\"ur Nanotechnologie, Forschungszentrum Karlsruhe, 76021 Karlsruhe,
Germany
\\
$^2$Institut f\"ur Theorie der Kondensierten Materie,
Universit\"at Karlsruhe, 76128 Karlsruhe, Germany}

\runninghead{I.~V.~Gornyi and A.~D.~Mirlin}{From quantum disorder to
quantum chaos} 
\begin{document}
\newcommand{\be}{\begin{equation}}
\newcommand{\ee}{\end{equation}}
\newcommand{\bea}{\begin{eqnarray}}
\newcommand{\eea}{\end{eqnarray}}
\newcommand{\br}{{\bf r}}
\newcommand{\bk}{{\bf k}}
\newcommand{\bq}{{\bf q}}
\newcommand{\bn}{{\bf n}}

\begin{abstract}

\vspace*{-1cm}

We study the statistics of wave functions in a ballistic chaotic
system. The statistical ensemble is generated by
adding weak smooth random potential, which allows us to apply the ballistic
$\sigma$-model approach. 
We analyze
conditions of applicability of the 
$\sigma$-model,
emphasizing the role played by the single-particle
mean free path and the Lyapunov exponent 
due to the
random potential.
In particular, we present a
resolution of the puzzle of repetitions of periodic orbits
counted differently by the
$\sigma$-model and by the trace formula. 

\end{abstract}

\maketitle


\section{INTRODUCTION}
\label{s1}

The central problem in the field of quantum chaos is
understanding of statistics of eigenfunctions and energy levels
of a quantum system whose classical counterpart is
chaotic and their relation to the underlying classical dynamics.
The Bohigas-Giannoni-Schmit \cite{bohigas84} conjecture states
that, generically, statistical properties of levels in such a system are
described (in the leading approximation) by universal results of the
random matrix theory (RMT). A related hypothesis concerning 
statistics
of wave functions has been put forward by Berry
\cite{Berry77} (see also \cite{sred}) who conjectured that an
eigenfunction of a chaotic billiard can be represented as a
random superposition of plane waves with a fixed absolute value
$k$ of the wave vector (determined by the energy $k^2/2m=E$,
where $m$ is the mass and we set $\hbar=1$). This implies
Gaussian statistics of the eigenfunction amplitude $\psi(\br)$,
\begin{equation}
{\cal P} \{ \psi \} \propto
\exp\left[-{\beta \over 2} \! \int d^2\br d^2\br'
\psi^*(\br)C^{-1}(\br,\br')\psi(\br')\right],
\label{Ppsi}
\end{equation}
determined solely by the correlation function
(we consider a two-dimensional system)
$C(\br_1,\br_2)\equiv\langle\psi^*(\br_1) \psi(\br_2) \rangle=
J_0(k|\br_1-\br_2|)/V$.
Here $\beta=1$ (or $2$) for a system with preserved (respectively
broken) time reversal symmetry, $V$ is the system area, and
$J_0$ the Bessel function.  Note that, when taken literally,
Eq.~(\ref{Ppsi}) contradicts the wave function normalization,
\be
\int d^2\br \left[\langle|\psi^2(\br) \psi^2(\br')| \rangle-
\langle|\psi^2(\br)|\rangle \langle|\psi^2(\br')| \rangle
\right]=0,
\label{norm}
\ee
since the integrand is equal to $C^2(\br,\br')>0$ according to
(\ref{Ppsi}). Therefore, the limits of validity of this
conjecture have to be understood.

Despite much effort spent in this direction, no proof of these
conjectures has been obtained via semiclassical methods, and
deviations from universality have not been calculated in a
controlled way. This is because the standard semiclassical tool
-- representation of a Green's function in terms of a sum over
classical trajectories -- is only justified for times much
shorter than the Heisenberg time $t_H=2\pi\hbar/\Delta$ (where
$\Delta$ is the mean level spacing), which is not sufficient for
the problems considered.

On the other hand, a considerable progress has been achieved in
investigation of the statistical properties of energy levels and
wave functions of diffusive disordered systems. In this case the
supersymmetry method \cite{efetov-book} serves as a tool for
a systematic
analytical
description of the level and eigenfunction
statistics. After averaging over an ensemble of realizations of the
random potential the problem is mapped onto a supermatrix
$\sigma$-model, which is further studied by various analytical
means. This has allowed one not only to prove the relevance of the
RMT results, but also to calculate system-specific deviations from
universality determined by the diffusive classical dynamics, see
\cite{PhysRep} for a review.

This success of the diffusive $\sigma$-model gave rise to an
expectation that the supersymmetry method may be also useful in the
context of quantum chaos. In a seminal paper \cite{MK}, 
Muzykantskii and Khmelnitskii
conjectured that a chaotic ballistic system can be described by a
ballistic $\sigma$-model, with the Liouville operator replacing the
diffusion operator in the action. They presented a derivation of this
model using the averaging over a white-noise disorder and
conjectured that it remains valid in the limit of vanishing
disorder. Subsequently, Andreev, Agam, Simons and Altshuler
\cite{AASA} proposed another derivation of the model, by
considering a clean system and employing the energy averaging
only. This led them to the conclusion that the statistical
properties of a chaotic system can be obtained from the results
found for diffusive systems \cite{PhysRep} by replacing
eigenvalues and eigenfunctions of the diffusion operators by
those of the (properly regularized) Liouville operator.

However, soon after publication of \ \cite{MK,AASA} these conclusions
were criticized from several points of view. Prange \cite{prange97}
showed that the energy averaging is insufficient to detect
non-universal corrections to the level statistics, in view of the statistical
noise. This indicates that one
cannot study an individual system but should rather consider some
ensemble of systems. The necessity of an ensemble
averaging has been also pointed out by Zirnbauer \cite{Zirn} who
concentrated on rigorous formulation and justification
of the Bohigas-Giannoni-Schmit conjecture of universality.
Bogomolny and Keating \cite{bogomolny96} emphasized the
discrepancy between the prediction of the ballistic
$\sigma$-model for the smooth part of the level correlation
function and the result of the diagonal approximation to the
semiclassical trace formula. Specifically, repetitions of
periodic orbits are counted differently within the two
approaches. These two issues (ensemble averaging and repetitions)
point to serious problems with the ballistic $\sigma$-model
approach and require clarification. This will be done below.

\section{BALLISTIC $\sigma$-MODEL FROM AVERAGING OVER SMOOTH
DISORDER}
\label{s2}

Let us come back to the derivation of the $\sigma$-model. It turns out
that the derivations proposed in both Refs.~\cite{MK} and \cite{AASA}
are, in fact, not quite correct. Specifically, if averaging over a
white-noise disorder with a mean free path $l$ is used (as in
\cite{MK}), the ballistic $\sigma$-model is only valid for momenta
$q\ll l^{-1}$
, i.e. in the region of validity of the
diffusive $\sigma$-model, but not in the ballistic range of
larger momenta \cite{BMM2}. On the other hand, the energy
averaging of Ref.~\cite{AASA} leaves a continuum of zero modes
allowing for arbitrary fluctuations of the supermatrix field
transverse to the energy shell and spoiling the derivation of the
ballistic $\sigma$-model, which requires
that
these fluctuations
be frozen \cite{Zirn}.

We note that the methods of \cite{MK} and \cite{AASA} may be
considered
as opposite extremes: while in Ref.~\cite{MK} the averaging over
a random potential with zero correlation length $d=0$  was
proposed, the energy averaging of Ref.~\cite{AASA} corresponds to
a random potential with $d=\infty$. As often happens, the truth
lies in between: the ballistic $\sigma$-model can be obtained if
one averages over a smooth random potential with a finite
correlation length $d$. In fact, this type of derivation had been
performed for the first time by W\"olfle and Bhatt
\cite{woelfle84} ten years before the notion of the ballistic
$\sigma$-model was introduced. However, the aim of
Ref.~\cite{woelfle84} was demonstration of the applicability of
the diffusive $\sigma$-model to the problem of a smooth random
potential, and the ballistic action appeared only implicitly as
an intermediate step of the calculation. This derivation was
generalized to the case of a random magnetic field in \cite{amw},
yet again the main point of that work was the diffusive action on
distances exceeding the transport mean free path. More recently,
it was emphasized \cite{TSE,BMM2} that averaging over a smooth
disorder is exactly the proper way of derivation of the ballistic
$\sigma$-model. This established the connection between
Refs.~\cite{woelfle84,amw} on the one hand, and
Refs.~\cite{MK,AASA} on the other hand.

Starting from a system with a Hamiltonian $\hat{H}$, we generate a
statistical ensemble \cite{BMM2} by adding a random potential
$U({\bf r})$ characterized  by a correlation
function $W(\br-\br')=\langle U(\br)U(\br')\rangle$
with a correlation length $d$.
Parameters of this random potential are assumed to
satisfy $k^{-1} \ll d \ll l_s \ll l_{\rm tr}$,
where $l_s =v\tau_s$, $l_{\rm tr}=v\tau_{\rm tr}$,
$v$ is the velocity, and $\tau_s (\tau_{\rm tr})$ is the single-particle
(respectively transport) relaxation time. Note that the potential
is smooth, $kd \gg 1$, so that $l_{\rm tr}/l_s \sim (kd)^2\gg 1$. In
order to apply this idea to a ballistic system of a
characteristic size $L$, we require $l_s \ll L\ll  l_{\rm tr}$.  The
condition $l_{\rm tr}\gg L$ preserves the ballistic nature of the
system, while the inequality $l_s \ll L$ guarantees that the
ensemble of quantum systems is large enough to produce meaningful
result.

After the ensemble averaging, the problem can be reduced
\cite{BMM2,woelfle84,amw,TSE}
to a ``non-local ballistic $\sigma$-model'' of a supermatrix field
$Q(\br,\bn)$ with the action (for definiteness, we consider the
case $\beta=2$)
\begin{eqnarray}
S[Q]&=&{\rm Str}\ln\left[E-{\hat H}+\omega\Lambda
-{i\over 2}\int \! d\bn' Q(\br,\bn')w(\bn,\bn')\right]\nonumber
\\
&-&\frac{\pi\nu}{4}\int d^2\br d\bn d\bn'  {\rm Str} \ Q(\br,\bn)
w(\bn,\bn')Q(\br,\bn'),
\label{SQ}
\end{eqnarray}
$\nu$ is the density of states,
$w(\bn,\bn')=2\pi\nu W(k|\bn-\bn'|)$ is the scattering
cross-section for the random potential, and $\bn$ is a unit
vector characterizing the direction of velocity on the energy
surface. Note that despite the non-local $\rm{Str}\ln$ form, the
action (\ref{SQ}) is not an exact representation of the original
problem, but rather a low-energy theory with only soft modes
kept: the momentum variable of $Q$ is constrained to the energy
surface, and $Q$ satisfies the usual $\sigma$-model condition
$Q^2=1$. To obtain the ballistic $\sigma$-model in the local form
(as proposed in \cite{MK,AASA}), one has to perform a gradient-
and frequency-expansion, which is justified provided $ql_s\ll 1$,
$\omega\tau_s\ll 1$. The result is \cite{TSE,BMM2}
\begin{eqnarray}
\label{wf6}
S[Q] &=& \pi\nu  \int d^2\br d\bn \,{\rm Str}\, \left[\Lambda
T^{-1}(\br,\bn ){\hat{\cal L}} T(\br,\bn )-{i\omega\over
2}\Lambda Q(\br ,\bn )\right]
 \nonumber\\
& + & {\pi\nu\over 4} \int d^2\br d\bn d\bn' \,{\rm Str}\,
Q(\br ,\bn )w(\bn ,\bn')Q(\br,\bn'),
\end{eqnarray}
where $Q(\br ,\bn)=T(\br ,\bn)\Lambda T^{-1}(\br ,\bn)$.
The symbol ${\hat{\cal L}}$ denotes the Liouville operator
characterizing the classical
motion; for a billiard ${\hat{\cal L}}=v\bn {\bf \nabla}$
supplemented
by appropriate boundary conditions \cite{WZ}.
The local ballistic $\sigma$-model (\ref{wf6})
is thus only applicable
on length
scales $\gg l_s$. From the
point of view of the semiclassical (periodic-orbit) theory, this
corresponds to the condition of applicability of the diagonal
approximation. On shorter scales, one has to use the more 
general, non-local, form (\ref{SQ}).

\section{EIGENFUNCTION STATISTICS IN A BALLISTIC SYSTEM}
\label{s3}

The two-point correlation function of the wave function
intensities is expressed in this approach as \cite{PhysRep,BMM2}
\begin{equation}
\langle|\psi^2(\br) \psi^2(\br')|\rangle
=
\lim_{\eta \to 0}\frac{\eta\Delta}{\pi}
\langle [G_{11}(\br,\br)G_{22}(\br',\br')
+
G_{12}(\br,\br')
G_{21}(\br',\br)] \rangle_{S[Q]},
\label{C2fG}
\end{equation}
where 
${\hat G}$ is
the Green's function in the field $Q$, \begin{equation}
{\hat G}=\left[E-{\hat H}+i\eta\Lambda-{i\over 2}\int d\bn'
Q(\br,\bn')w(\bn,\bn')\right]^{-1},
\label{fG}
\end{equation}
and the subscripts $1,2$ refer to the advanced-retarded
decomposition (the boson-boson components being implied).
Here $\eta$ is an infinitesimal positive, and $\omega$ in the
action $S[Q]$ is given by $\omega=i\eta$.

We first evaluate Eq.~(\ref{C2fG}) in the zero-mode approximation,
$Q(\br)=Q_0$. The Green's function (\ref{fG}) is given in the
leading order by
\bea
G_0(\br,\br')&=& i {\rm Im} G_R(\br,\br')Q_0+
{\rm Re} G_R(\br,\br'),
\label{G0} \\
G_R(\br,\br')&=&\langle \br | (E-{\hat H} +
i/{2\tau_s})^{-1} |\br'\rangle.
\label{GRA}
\eea
Substituting Eq.~(\ref{G0}) in (\ref{C2fG}) and expanding the
action (\ref{SQ}) up to the linear-in-$\eta$ term,
$S[Q]\simeq
\pi\nu\eta V {\rm Str} Q_0 \Lambda $, one finds, in a
full analogy with the case of diffusive systems \cite{Prigodin},
\bea
&&V^2\langle|\psi^2(\br_1) \psi^2(\br_2)|\rangle
\simeq 1+k_q(\br_1,\br_2); \label{ZM0} \\
&&k_q(\br,\br')={\rm Im}G_R(\br,\br')
{\rm Im}G_R(\br',\br)/(\pi\nu)^2,
\label{kq}
\eea
with the two contributions on the r.h.s. of (\ref{ZM0})
originating from the terms $\langle G_{11}G_{22} \rangle $ and
$\langle G_{12}G_{21}\rangle$ in (\ref{C2fG}), respectively.
The result (\ref{ZM0}), corresponding exactly to the conjecture
(\ref{Ppsi}) of the Gaussian statistics, is in
conflict with the wave function normalization, as explained
above.

To resolve this problem, we evaluate the term
$\langle G_{11}G_{22} \rangle $
more accurately by expanding the Green's function
(\ref{fG}) to the order $\eta$ and the action (\ref{SQ}) to the
order $\eta^2$. While these terms (usually neglected in the
$\sigma$-model calculations) are of the next order in $\eta$
and may be naively thought to vanish in the limit $\eta\to 0$
performed in (\ref{C2fG}), this is not so, since $Q_0 \propto
\eta^{-1}$.
As a result, we get in the zero-mode
approximation
\be
V^2 \langle|\psi^2(\br_1) \psi^2(\br_2)|\rangle_{\rm ZM}-1
= k_q(\br_1,\br_2)-{\bar k}_q(\br_1)-
{\bar k}_q(\br_2)+{\bar{\bar k}}_q
\label{C2zeromode}
\ee
(terms of still higher orders in $\eta$ produce
corrections small in the parameter $\Delta\tau_s\ll 1$), where
${\bar k}_q(\br)= V^{-1}\int  d^2\br' k_q(\br,\br'),
\quad
{\bar {\bar k}}_q= V^{-2}\int  d^2\br d^2\br' k_q(\br,\br').
\label{kbar}
$

The contribution of non-zero modes
is expressed in terms of the $\sigma$-model propagator,
\cite{BMM2}
\be
V^2\langle|\psi^2(\br_1) \psi^2(\br_2)|\rangle_{\rm
NZM}={\tilde \Pi}_B(\br_1,\br_2),
\label{tildeP}
\ee
where ${\tilde \Pi}_B(\br_1,\br_2)={\Pi}_B(\br_1,\br_2)-
{\Pi}^{(0)}_B(\br _1,\br_2)$ describes the (integrated over
direction of velocity) probability of classical propagation from
$\br_1$ to $\br_2$,
\begin{eqnarray}
&&\Pi_B(\br_1,\br_2)=\int\int d\bn_1
d\bn_2{\cal D}(\br_1\bn_1,\br_2\bn_2),\nonumber \\
&&{\hat{\cal L}}{\cal D}=(\pi\nu)^{-1}\left[\delta(\br_1-\br_2)
\delta(\bn_1-\bn_2)-V^{-1} \right],
\label{PB}
\end{eqnarray}
with the contribution ${\Pi}^0_B(\br _1,\br_2)$ of direct
propagation (before the first event of disorder scattering)
excluded \cite{footnote}.

We analyze now the total result given by the sum of
(\ref{C2zeromode}) and (\ref{tildeP}). First of all, we
stress that it satisfies exactly the condition (\ref{norm}) of
wave function normalization.
Next, we consider sufficiently short distances,
$|\br_1-\br_2|\ll l_s$. In this case the correlation function is
dominated by the first term in the r.h.s. of
Eq.~(\ref{C2zeromode}), returning us to the result (\ref{ZM0}).
Furthermore, we can generalize this result to higher correlation
functions,
\bea
&&\langle\psi^*(\br_1)
\psi(\br'_1) \dots  \psi^*(\br_n)
\psi(\br'_n)\rangle =-\frac{1}{2V(n-1)!}\nonumber \\
&&\times\lim_{\eta\to 0}(2\pi\nu\eta)^{n-1}
\left\langle\sum_\sigma \prod_{i=1}^n{1 \over \pi \nu}
G_{p_i{p_{\sigma(i)}}}(\br_i,\br'_{\sigma(i)})
\right\rangle_{S[Q]},
\label{eq451}
\eea
where the summation goes over all
permutations $\sigma$ of the set $\{1,2,\dots,n\}$,
$p_i=1$ for $i=1,\dots,n-1$, and $p_n=2$.
If all the points $\br_i,\br_i^\prime$ are within a distance
$\ll l_s$ from each other, the leading contribution to this
correlation function is given by the zero-mode approximation with
higher-order terms in $\eta$ neglected [i.e. by the same
approximation which leads to Eq.~(\ref{ZM0})], yielding
\bea
&&V^n\langle\psi^*(\br_1)
\psi(\br'_1) \dots  \psi^*(\br_n)
\psi(\br'_n)\rangle=\sum_\sigma\prod_{i=1}^{n}
f_F(\br_i,\br_{\sigma(i)}^\prime),\nonumber \\
\label{higher}
&&f_F(\br,\br')=-{\rm Im}G_R(\br,\br')/(\pi\nu).
\label{fF}
\eea
This result
is identical to the statement
of the Gaussian statistics of eigenfunctions conjectured in
\cite{Berry77,sred}. We have thus proven that
Eq.~(\ref{Ppsi})
holds within a spatial region of an extension
$\ll l_s$, with the kernel $C(\br_1,\br_2)=
f_F(\br_1,\br_2)/V$ given by
Eq.~(\ref{fF}).

We turn now to the behavior of the correlator
$\langle|\psi^2(\br_1) \psi^2(\br_2)|\rangle$
at larger separations $|\br_1-\br_2|\gg l_s$.
In this situation, the correlations are dominated by the
contribution (\ref{tildeP}) of non-zero modes. Let us further
note that the smooth part of the zero-mode contribution
(\ref{C2zeromode}) (i.e. with Friedel-type oscillations
neglected) is
exactly equal to ${\Pi}^{(0)}_B$. Therefore, the smoothed
correlation function is given by the classical propagator,
\be
V^2\langle|\psi^2(\br_1) \psi^2(\br_2)|\rangle_{\rm smooth}-1=
\Pi_B(\br_1,\br_2),
\label{smooth}
\ee
independent of the relation between $|\br_1-\br_2|$ and $ l_s$.
The mean free path $l_s$ manifests itself only in setting the
scale on which the oscillatory part of
$\langle|\psi^2(\br_1) \psi^2(\br_2)|\rangle$ gets damped.

To summarize, the disorder averaging generates the scale $l_s$
which separates the regions of applicability of the Gaussian
statistics and of the quasiclassical theory (\ref{wf6}).
However, this is not the full story yet.
As we demonstrate below, the disorder averaging induces one
more scale which plays a crucial role for the problem considered.

\section{PROBLEM OF REPETITIONS}
\label{s4}

As was mentioned in Sec.~\ref{s1}, Ref.~\cite{bogomolny96}
emphasized discrepancy between the $\sigma$-model and the trace
formula as concerned the counting of repetitions in the
expression for the
level correlation function. As we demonstrate in this Section,
this problem, while not affecting (except for a
non-generic situation when the points $\br_1$ and $\br_2$ lie
close to one and the same  short periodic orbit) the lowest-order
correlation function
$\langle|\psi^2(\br_1)\psi^2(\br_2)|^2\rangle$ considered in
Sec.~\ref{s3}, is also relevant to higher-order correlators of
wave function amplitudes. We will clarify the origin of this
puzzling discrepancy and formulate the conditions under which
each of the two results apply. We
assume
$\beta=1$ in this section (so that eigenfunctions are real) and
start our consideration from the correlation function
\be
\label{e4.1}
\Gamma^{(4)}(\br-\br')=\langle \psi_\mu(\br)\psi_\nu(\br)
\psi_\rho(\br)\psi_\sigma(\br)\psi_\mu(\br')\psi_\nu(\br')
\psi_\rho(\br')\psi_\sigma(\br')\rangle,
\ee
where $\psi_{\mu,\nu,\rho,\sigma}$ are different eigenfunctions with
sufficiently close energies. We will further assume that
$l_s\ll|\br-\br'|\ll l_{\rm tr}$, i.e. the distance $|\br-\br'|$ is in the
expected range of applicability of the ballistic $\sigma$-model (\ref{wf6}). 
The objects of the type (\ref{e4.1}) naturally arise when one studies
fluctuations of matrix elements of the electron-electron interaction,
which are important for statistics of electron transport through
quantum dots, see \cite{blanter,AGKL}.

We first consider the situation of a sufficiently large system size
$L$ ({\it e.g.}, we can assume a diffusive system, $L\gg l_{\rm tr}$); in
this case the value of $L$ will be irrelevant for the results (except
for normalization of wave functions). In the end of the section 
we will generalize our
conclusions to the case of a ballistic system.
We start by performing an averaging over an auxiliary random
potential with a single-particle
mean free path $\tilde{l}_s=v\tilde{\tau}_s$ satisfying
$|\br-\br'|\ll\tilde{l}_s\ll L$ (which is much weaker than our ``main''
random potential and will thus not enter the final result). This auxiliary
averaging allows us to present (\ref{e4.1}) in the form
\be
\label{e4.2}
\Gamma^{(4)}(\br-\br')=(\pi\nu)^{-4}\langle[{\rm
Im}\,G_R(\br,\br')]^4\rangle = {3\over 8}(\pi\nu)^{-4}\langle
G_R^2(\br,\br')G_A^2(\br,\br') \rangle ,
\ee
where $G_{R,A}=(E-\hat{H}\pm i/2\tilde{\tau}_s)^{-1}$ is a
Green's function in a given realization of the ``main'' random
potential, the averaging over which remains to be performed in
the r.h.s. of (\ref{e4.2}), as indicated by $\langle \dots \rangle$.
The
cumulant $\langle\langle \psi^4(\br)\psi^4(\br')\rangle\rangle$
of a single eigenfunction can be also reduced to the form
(\ref{e4.2}), if one uses the $\sigma$-model approach to average
over the auxiliary random potential, in full analogy with the
derivation of Eqs.~(\ref{ZM0}), (\ref{kq}) (in view of
$|\br-\br'|\ll \tilde{l}_s$ the zero-mode approximation is
appropriate). An analogous trick of an auxiliary averaging was
used in \cite{gm01} in order to study ballistic wave function
correlations in a random magnetic field.

Let us first evaluate the r.h.s. of (\ref{e4.2}) using the ballistic
$\sigma$-model. The calculation is trivial and yields
\be
\label{e4.3}
(2\pi^2\nu^2)^{-2}\langle
G_R^2(\br,\br')G_A^2(\br,\br')\rangle=2\Pi_B^2(\br,\br') \simeq
{2\over (\pi k|\br-\br'|)^2}.
\ee
The result (\ref{e4.3}) is fully transparent from the point of view
of diagrammatics: there are two possibilities to couple the
retarded and advanced Green's functions in two
``ballistic-diffuson'' ladders, yielding the factor of 2 in front
of the squared propagator $\Pi_B^2(\br,\br')$.

We are going to show now that the result (\ref{e4.3}) of the ballistic
$\sigma$-model is only correct for sufficiently large distances,
$|\br-\br'|\gg l_L$ (the scale $l_L$ will be specified below), while in
the opposite limit, $|\br-\br'|\ll l_L$, it is wrong by factor of 2.
To evaluate the r.h.s. of (\ref{e4.2}), we use the path integral
approach \cite{MAW}.
The product of the four Green's functions in (\ref{e4.2})
can be written as (we set $\br=0$ and ${\bf R}=\br'-\br$)
\begin{eqnarray}
&&\langle G_R^2(0,{\bf R})G_A^2(0,{\bf R})\rangle =
\prod_{i=1}^4\int_0^\infty \! \!\! dt_i
\int_{\br_i(0)=0}^{\br_i(t_i)={\bf R}}\! \! \! \! {\cal
D}\br_i \quad \exp[iS_{\rm kin}-S_{\rm W}]
\nonumber \\  &&\times
\exp[iE(t_1+t_2-t_3-t_4)-(t_1+t_2+t_3+t_4)/{\tilde \tau}_s];
\label{PI}\\
&&S_{\rm kin}={m\over 2}\left(\int_0^{t_1}dt {\dot \br}_1^2+
\int_0^{t_2}dt {\dot \br}_2^2 -\int_0^{t_3}dt {\dot \br}_3^2-
\int_0^{t_4}dt {\dot \br}_4^2\right);\nonumber \\
&&S_{\rm W}={1\over 2}(S_{11}+S_{22}+S_{33}+S_{44})+
S_{12}+S_{34}-
S_{13}-S_{14}-S_{23}-S_{24};\nonumber \\
&&S_{ij}=\int_0^{t_i}\int_0^{t_j}
W(\br_i(t)-\br_j(t'))dt dt',
\label{GGGG}
\end{eqnarray}
where the paths $\br_1(t_1),\br_2(t_2)$ correspond to
the retarded, and $\br_3(t_3),\br_4(t_4)$ to the advanced Green's
functions. It is useful to perform the change
of variables, introducing
$T=(t_1+t_2+t_3+t_4)/4, \
t_{-}=(t_1+t_2-t_3-t_4)/2, \
\tau_{12}=t_1-t_2, \
\tau_{34}=t_3-t_4,$
and
$
{\bf R_+}=(\br_1+\br_2+\br_3+\br_4)/4, \
{\bf R_-}=(\br_1+\br_2-\br_3-\br_4)/2, \
\br_{12}=\br_1-\br_2, \
\br_{34}=\br_3-\br_4.
$
The kinetic part of the action
then reads
\begin{eqnarray}
S_{\rm kin}&\simeq& m\int_0^T dt\left\{
\left(-{t_{-}\over
T}+\frac{\tau_{12}^2-\tau_{34}^2}{4T^2}\right)
{\bf {\dot R}_+}^2
\right.
\nonumber \\
&+&
\left.
2 {\bf {\dot R}_+}{\bf {\dot R}_-}+\frac{{\dot \br_{12}}^2-
{\dot \br_{34}}^2}{4}-{\bf {\dot R}_+}
\frac{\tau_{12}{\dot \br_{12}}-\tau_{34}{\dot\br_{34}}}{2T}
\right\}.
\end{eqnarray}
Since we are interested in the ballistic scales ($\ll
l_{\rm tr}$) it is convenient to split $\br,\br_{12},\br_{34}$
into components parallel ($||$) and perpendicular
($\perp$) to ${\bf R}$.
As shown below, $S_{\rm W}$ depends only on the $\perp$-components.
Shifting the
parallel components of $\br_{12},\br_{34}$ via
$\rho_{12}=r_{12}^{||}-R_+^{||}\tau_{12}/T$, 
$\rho_{34}=r_{34}^{||}-R_+^{||}\tau_{12}/T$,
we can perform integration over
$\rho_{12},\rho_{34},\tau_{12}$, and $\tau_{34}$,
which yields the factor $(T/R)^2$.
Furthermore, integration over $R_+^{||}, R_-^{||},t_-$, 
and $T$ produces the factor $m/2E$ and sets $T=R/v$ 
with $v=(2E/m)^{1/2}$.
Finally, introducing
\begin{equation}
\rho_{\pm}=(r_{12}^{\perp}\pm r_{34}^{\perp})/2,\quad
r_{\pm}=R_{\pm}^{\perp},
\end{equation}
we simplify the action to the form
\begin{equation}
{\tilde S}_{\rm kin}=m\int_0^T dt
\left\{2{\dot r}_+{\dot r}_-
+{\dot \rho_{+}}{\dot \rho_{-}}\right\}.
\label{SkinFin}
\end{equation}

Now we turn to the disorder-induced part of the action,
$S_{\rm W}$ (where all $t_i$ can be approximated by T, see \ \cite{MAW}).
Expanding the correlator $W$
up to the second order in $\br$, we obtain
\be
S_{\rm W}\simeq
\int_0^T\!\!{\cal U}(\rho_{-}(t),\rho_{+}(t))dt
-2\int_0^T\!
\{G(\rho_{-}(t))+G(\rho_{+}(t))\}r_-^2(t)dt,
\label{SWFG}
\ee
where
\be
{\cal U}(y,y')=
2\{F(y)+F(y')\}-F(y+y')-F(y-y').
\label{Urr}
\ee
Here the functions $F(y)$ and $G(y)$ are given by
\be
F(y)\equiv\int_0^\infty{dx\over v}[W(x,0)-W(x,y)],
\quad
G(y)\equiv\int_0^\infty{dx\over v}
{\partial^2\over \partial y^2} W(x,y),
\label{defG}
\ee
with the following asymptotic
behavior
$F(y\ll d)\simeq -G(0)y^2/2$, $F(y\gg d)\simeq
\tau_s^{-1}$, $G(0)=-m^2v^2/\tau_{\rm tr}$, and $G(y\gg d)\to 0$.

We examine first the contribution of the region
$\rho_{+},\rho_{-}\gg d$, where all four paths
are uncorrelated. Then the action (\ref{SWFG}) is
large, $S_{\rm W}\simeq 2T F(y\gg d)=2T/\tau_s$, since
the phases acquired by the waves
traveling along these paths do not cancel each other.
Thus the contribution of this region
decays exponentially at $R\gg l_s$ (which is the scale of our
interest), in the same way as the single Green's
function does. We conclude that the path integral (\ref{PI}) is
dominated by paths with at least one of $\rho_{\pm}$ much less
than $d$, which corresponds to correlations between ``retarded''
and ``advanced'' paths.

Let us now consider the contribution of
$\rho_{+}\gg d, \ \rho_{-}\ll d$, when the two
``ballistic diffusons'' (formed by the pairs of paths $1,3$ and
$2,4$) are not correlated. Then $S_{\rm W}$ acquires a form
\begin{equation}
S_{\rm W}\simeq\int_0^T
\left[{{2w_0}\over v}r_{-}^2(t)+
{{w_0}\over v}\rho_{-}^2(t)\right]dt,
\label{SWdiff}
\end{equation}
where $w_0=-vG(0)
\sim
v(d^2\tau_s)^{-1}$. Combining Eqs.~(\ref{SkinFin}) and
(\ref{SWdiff}) we see that the variables $r_{\pm}$ and
$\rho_{\pm}$  separate, each of the two pairs
being characterized by the diffusive action \cite{MAW}
(describing the dynamics of the ``center of mass'' $r_+$ and
the distance between the two ``diffusons'' $\rho_+$,
respectively). This gives squared propagator $\Pi_B^2(\br,\br')$,
as in Eq.~(\ref{e4.3}). The region $\rho_{-}\gg d, \ \rho_{+}\ll
d$ (diffusons formed by paths $1,4$ and $2,3$) yields an
identical contribution, reproducing the factor 2 in the
$\sigma$-model result (\ref{e4.3}).

However, the approximation (\ref{SWdiff}) is only justified at
sufficiently large $T$. Indeed, the action (\ref{SkinFin}),
(\ref{SWdiff}) implies a diffusion of the velocity
$\dot{\rho}_{+}/v$ with the diffusion coefficient $1/\tau_{\rm tr}$.
Taking into account the boundary conditions
$\rho_+(0)=\rho_+(T)=0$, it requires a time of the order of
\be
\label{taul}
\tau_L\sim\tau_{\rm tr}(d/l_{\rm tr})^{2/3}
\ee
for $\rho_+$ to reach a value $\sim d$, which implies
$T\gg \tau_L$ as a condition of validity of Eq.~(\ref{SWdiff}) and
thus of the result (\ref{e4.3}). As we discuss below, the scale $\tau_L$ 
has a
meaning of the inverse Lyapunov exponent; hence the subscript
``L''.

At shorter distances $R\ll l_L=v\tau_L$, both $\rho_{+}$ and
$\rho_{-}$ are small compared to $d$. 
Expanding
(\ref{SWFG}) up to the second
order in both $\rho_{+}$ and $\rho_{-}$ we have
\begin{equation}
S_{\rm W}\simeq\int_0^T
\left[{{4w_0}\over v}r_{-}^2(t)+
{{w_2}\over v}\rho_{+}^2(t)\rho_{-}^2(t)\right]dt,
\label{SWfin}
\end{equation}
where $w_2=-vG''(0)/2
\sim
w_0/d^2$ and we neglected cross-terms of the type
$r_{\perp}^2\rho_{\pm}^2$, which are small compared to the
first term in the integrand in (\ref{SWfin}).
We note that the variables $r_{\pm}$ and $\rho_{\pm}$ 
again separate. The first pair corresponds to the
diffusion of ${\dot r}_+$ [with a diffusion coefficient
two times smaller than in Eq.~(\ref{SWdiff})], while the second
one now describes the divergence of the two pairs of paths. The
differential equation for a Green's function which corresponds to
the $\rho$-part of the action reads
\begin{equation}
\left({\partial \over {\partial t}}
-{i\over {m}}
\frac{\partial^2}{\partial\rho_{+}\partial\rho_{-}}+
\frac{w_2}{v}\rho_{+}^2\rho_{-}^2
\right)g(\rho_{+},\rho_{-},t)=
\delta(t)\delta(\rho_{+})\delta(\rho_{-}),
\label{diffur}
\end{equation}
and we are interested in the quantity $g(0,0,t)$, the correlation
function (\ref{e4.3}) being given by
\be
\label{e4.51}
(2\pi^2\nu^2)^{-2}\langle
G_R^2(\br,\br')G_A^2(\br,\br')\rangle={2\over\pi m^3v^3R}g(0,0,R/v)\ .
\ee

Now we analyze the solution of (\ref{diffur}). Clearly,
$\tau_L\equiv(m^2v/w_2)^{1/3}$ sets a characteristic time scale for
this equation.
It is remarkable that upon the Fourier transformation,
$\rho_+\rightarrow i(mv)^{-1}\partial/\partial\phi$ and
$\partial/\partial\rho_{+} \rightarrow -imv\phi$,
the operator in the l.h.s. of (\ref{diffur}) coincides
with the one obtained by Aleiner and Larkin in Ref.\cite{Alla}
from the analysis of divergence of {\it classical} paths in a
weak smooth
random potential. As shown in \cite{Alla}, the scale $\tau_L$ has the
meaning of a corresponding inverse Lyapunov exponent.
We have thus demonstrated that the
same scale arises in the treatment of a {\it quantum} ($l_s\gg
d$) random potential, despite the diffractive nature of the
scattering \cite{tautr}.

For $t\ll \tau_L$ one can neglect the term  $\rho_{+}^2\rho_{-}^2$ in the
l.h.s of (\ref{diffur}), yielding
\be
\label{e4.52}
g(\rho_+,\rho_-,t)\simeq {m\over2\pi t}
\exp {i\rho_+\rho_-\over mt} \ ,\qquad t\ll \tau_L.
\ee
In fact, the solution remains a function of the product $\rho_+\rho_-$
for all $t$, and Eq.~(\ref{diffur}) can be reduced, in dimensionless
variables $\tau=-i^{5/3}t/\tau_L$,
$r=(i/4)^{-1/6}(m\rho_+\rho_-/\tau_L)^{1/2}$, to an equation for
a 2D anharmonic oscillator,
\be
\label{e4.53}
(-i\partial_\tau-r^{-1}\partial_r r\partial_r+r^4)
\tilde{g}(r,\tau)=\delta^2(r)\delta(\tau) .
\ee
Therefore, at  $\tau\gg 1$ (corresponding to $t\gg \tau_L$) its solution
decays exponentially, $g\propto \exp\{-i^{2/3}\epsilon_0t/\tau_L\}$, where
$\epsilon_0$ is the (dimensionless) oscillator ground-state energy.
However, as explained above, at this time scales the
approximation $\rho_+,\rho_-\ll d$ leading to Eq.~(\ref{diffur})
loses its validity, and the result is dominated by the
contributions with either $\rho_+\gg d$ or $\rho_-\gg d$
described by the action (\ref{SWdiff}) and yielding the result
(\ref{e4.3}).

We thus conclude that 
\be
\label{e4.54}
(2\pi^2\nu^2)^{-2}\langle
G_R^2(\br,\br')G_A^2(\br,\br')\rangle= {1\over(\pi k R)^2}
{\cal F}(R/l_L) ,
\ee
where ${\cal F}(x)$ is a parameterless function with asymptotics 
${\cal F}(x\ll 1)=1$ and ${\cal F}(x\gg 1)=2$. To calculate the
crossover function ${\cal F}(x)$, one has to solve a
differential equation\cite{Hikami}
\begin{equation}
\left({\partial \over {\partial t}}
-{i\over {m}}
\frac{\partial^2}{\partial\rho_{+}\partial\rho_{-}}+
{\cal U}(\rho_{+},\rho_{-})
\right)g(\rho_{+},\rho_{-},t)=
\delta(t)\delta(\rho_{+})\delta(\rho_{-}),
\label{eqUrr}
\end{equation}
with ${\cal U}(\rho_{+},\rho_{-})$ defined in (\ref{Urr}); the desired
correlation function is then given by Eq.~(\ref{e4.51}).
The solution in the crossover region depends on the
specific form of the random potential correlation
function $W(r)$ and, in general, can only be found numerically;
however, the explicit form of the function ${\cal F}(x)$ is not
important for our analysis. A similar analysis can be performed for
higher-order correlation function $\langle
G_R^n(\br,\br')G_A^n(\br,\br')\rangle$, yielding the crossover
from  ${\cal F}_{(n)}(x\ll 1)=1$ to ${\cal F}_{(n)}(x\gg 1)=n!$.

Therefore, the ballistic $\sigma$-model result is only valid
at distances $R\gg l_L$.
At shorter scales, the four paths do not split into two pairs
yielding each a ballistic diffuson (as the $\sigma$-model approach
assumes) but rather propagate together, remaining all strongly
correlated. In the diagrammatic language this means that all the four
Green's functions are coupled by impurity lines. This is closely
related to the non-Markovian
memory effects \cite{memory} showing up in transport in
a smooth random potential.

For ballistic systems this implies that the $\sigma$-model approach is
only valid provided the random potential over which the averaging is
performed is sufficiently strong, so that $l_L\ll L$. Note that such an
averaging can be characterized as ``strongly invasive'': the Lyapunov
exponent due to the random potential $\tau_L^{-1}$ is much larger
than that of the clean system itself (which is $\sim v/L$ for a generic
system of a size $L$). Therefore, although such a random potential
does not essentially influence (in view of $l_{\rm tr}\gg L$)
the positions of Ruelle resonances governing the relaxation rate
in the chaotic system, it strongly affects the Lyapunov
exponents.

In the opposite case $l_L\gg L$ (corresponding to a ``non-invasive''
averaging, keeping intact all the classical features of the system)
the ballistic $\sigma$-model does not give correct results for
non-universal level and eigenfunction correlations. More precisely, it
is still valid when applied to the lowest-order correlation function
$\langle|\psi^2(\br)\psi^2(\br')|\rangle$, see
Eq.~(\ref{smooth}). However, when one considers higher-order
correlation functions (or, in fact,
$\langle|\psi^2(\br)\psi^2(\br')|\rangle$ with $\br$ and $\br'$
located close to a short periodic orbit), the approach breaks down,
since it does not take into account correlations between four or more
close paths. This is precisely the problem which shows up in the
counting of repetitions in the formula for the level correlation
function. In that case, relevant paths for both  $G_R$ and $G_A$
wind $n$ times in the vicinity of a periodic orbit.
If $L\ll l_L$, all these paths are correlated (in the same
way as four Green's functions in the example considered above).
The $\sigma$-model neglects this fact,
combining the two paths (``R'' and ``A'') in a ladder, which can be done in $n$
possible ways.
This leads to an overestimate of the contribution by a factor of $n$ 
(similarly to the
overestimate of the correlation function (\ref{e4.3}) at
$R\ll l_L$ by a factor of 2). Therefore, the results of the
$\sigma$-model \cite{AASA} and of the trace formula
\cite{bogomolny96} for the smooth part of the level correlation
function correspond to different types of averaging: the former
to $l_L\ll L$, while the latter to $l_L\gg L$ (in both the cases
the condition $l_s\ll L$ should be assumed, see Sec.~\ref{s2}).

It is worth noting that we assumed the clean system to be
characterized by a single scale $L$, so that the corresponding
Lyapunov exponent is $\sim L^{-1}$. An example of a system which does
not belong to this class is given by a billiard with a surface
roughness producing a diffuse boundary scattering \cite{BMM2}.
In this case repetitions give a negligibly small
contribution, and the $\sigma$-model approach as applied in
\cite{BMM2} is justified with the only assumption $l_s\ll L$.

Finally, let us note a close similarity between the condition
$l_L\gtrsim L$ of the failure of the $\sigma$-model and the
condition $l_L\gtrsim R_c$ (with $R_c$ the cyclotron radius)
under which memory effects develop into adiabaticity which affects
strongly magnetoresistivity in a smooth disorder
\cite{magres}.
In both cases, correlations between multiple traversals of a
periodic (respectively cyclotron) orbit
lead to breakdown of the Boltzmann kinetic equation.

\section{CONCLUSIONS}
\label{s5}

We have studied the statistics of wave function in a chaotic system
on the ballistic scale. To define the statistical ensemble, we have
added to the system a smooth quantum random potential with a
correlation length $d\gg k^{-1}$.
We required the corresponding quantum
(single-particle) mean free path to satisfy  $l_s\ll L$, which
ensures that the
ensemble of quantum systems is sufficiently large and provides
meaningful statistics. Using the ballistic $\sigma$-model
approach, we have shown that on scales $\ll l_s$ the wave
function statistics is Gaussian, Eq.~(\ref{higher}),
proving the Berry's conjecture \cite{Berry77,sred}. On larger scales
Friedel-type oscillations in the correlation function
$\langle|\psi^2(\br)\psi^2(\br')|\rangle$ are exponentially damped,
and the latter is given by the classical ballistic propagator,
Eq.~(\ref{tildeP}). However, an attempt to use the $\sigma$-model
approach for calculation of higher-order correlation functions in this
regime makes us to face a problem: the $\sigma$-model does not take
into account strong correlations between 4 (or more) Green's functions
dominated by paths which are all close to each other. Considering the
correlation function (\ref{e4.2}) as an example and performing its
detailed analysis, we demonstrated that the ballistic $\sigma$-model
result is only correct on distances $R\gg l_L=v\tau_L$,
where $\tau_L^{-1}$ is the classical Lyapunov exponent associated
with the random potential, see Eq.~(\ref{taul}). It is remarkable
that the Lyapunov scale $l_L$ arises despite the fact that the
random potential is assumed to be quantum, with $l_s\gg d$.
On shorter distances, $R\ll l_L$, the $\sigma$-model result for
(\ref{e4.2}) is wrong by factor of two. In this regime the path
integral approach is more appropriate; it allows one also to
describe (at least, in principle) the crossover between the two
regimes ($R\sim l_L$).

A completely analogous situation is encountered when one studies the
level correlation function. Our results thus resolve the
discrepancy between the predictions of the $\sigma$-model \cite{AASA}
and of the trace formula \cite{bogomolny96} for the
smooth part of the two-level correlation function. The
$\sigma$-model result is generically valid provided $l_L\ll L$,
where $L$ is the system size. This condition means that the
averaging is ``invasive'', since it strongly affects the Lyapunov
exponent (although it does not affect, in view of the assumption
$l_{\rm tr}\gg L$, positions of the Ruelle resonances)\cite{WL}. In
the opposite case of a ``non-invasive'' (preserving all classical
parameters of the system) averaging, $l_L\gg L$, the
$\sigma$-model approach neglecting correlations between
multiple
traversals of a periodic orbit loses its validity,
and the smooth part of the level correlation function is given by
the trace formula \cite{bogomolny96}. Calculation of the
system-specific oscillatory contribution in this situation
remains an open problem. Such contributions are related to the
behavior of the spectral form-factor at times close to the
Heisenberg time $t_H$, which is beyond the region of validity of
the trace formula. Bogomolny and Keating \cite{bogomolny96}
proposed a procedure allowing to obtain the oscillatory
contribution from the trace formula. However, in view of the {\it
ad hoc} nature of their suggestion, the status of the result is
unclear. On the other hand, the $\sigma$-model approach, which
would be more appropriate for an analysis of contributions of
this kind (non-perturbative from the field-theoretical point of
view), does not treat properly correlations of multiple close
paths. A natural idea would be to combine the two approaches,
{\it i.e.} to formulate a field theory which would contain not
only conventional ``2-diffusons'' ({\it i.e.} those generated by
a product $G_RG_A$ of two Green's functions) ${\cal
D}(\br\bn,\br'\bn')$, but also ``4-diffusons'', ``6-diffusons'',
etc. This remains a challenge for future research.

\section*{ACKNOWLEDGMENTS}
We thank B.A.~Muzykantskii and D.G.~Polyakov for useful discussions.
This work was supported by the SFB195
and the Schwerpunktprogramm
``Quanten-Hall-Systeme'' der Deutschen
Forschungsgemeinschaft, the INTAS grant 99-1705,
and the RFBR grant 99-02-17110.


\begin{thebibliography}{99}
\bibitem{WZ} Note that in Ref. \cite{MK} the $\sigma$-model
action was presented in a different (Wess-Zumino) form
which makes explicit its invariance (characteristic for
$\sigma$-models) with respect to gauge transformation 
$T(\br,\bn)\to T(\br,\bn)U(\br,\bn)$ with $[U,\Lambda]=0$.
\bibitem{footnote} In fact, the operator ${\hat{\cal L}}$ in
Eq.~(\ref{PB}) contains an additional term describing scattering
by the random potential, $\delta{\hat{\cal L}}f(\bn)=\int
d\bn'[f(\bn)-f(\bn')]w(\bn,\bn')$. However, this term does not
affect essentially the propagator $\Pi_B(\br_1,\br_2)$ in view
of the assumption $l_{\rm tr}\gg L$.
\end{thebibliography}
\end{document}